\def\year{2019}\relax
\documentclass[letterpaper]{article} 
\usepackage{aaai19}  
\usepackage{times}  
\usepackage{helvet} 
\usepackage{courier}  
\usepackage[hyphens]{url}  
\usepackage{graphicx} 
\usepackage{color, colortbl}
\definecolor{LightCyan}{rgb}{0.88,1,1}
\usepackage{booktabs} 
\usepackage{amsmath}
\usepackage{mathtools}
\usepackage{url}
\usepackage{textcomp}

\usepackage{subfigure}
\usepackage{enumitem}
\usepackage{tabularx}
\usepackage{makecell}
\usepackage{multirow}
\usepackage{marvosym}
\usepackage{amsfonts}
\usepackage[normalem]{ulem}
\title{Organizational Artifacts of Code Development}

\author{ \Large \textbf{Parisa Kaghazgaran, Nichola Lubold, Fred Morstatter}\\ 
Amazon, Honeywell, University of Southern California\\ 

parisaka@amazon.com, nichola.lubold@honeywell.com, fredmors@isi.edu   
}
\newcommand{\squishlist}{
   \begin{list}{$\bullet$}
   { \setlength{\itemsep}{0pt}
       \setlength{\parsep}{1pt}
       \setlength{\topsep}{1pt}
       \setlength{\partopsep}{0pt}
       \setlength{\leftmargin}{1em} 
       \setlength{\labelwidth}{1em}
       \setlength{\labelsep}{0.5em}
    } }

\newcommand{\squishlisttwo}{
   \begin{list}{$\bullet$}
       { \setlength{\itemsep}{0pt}
           \setlength{\parsep}{0pt}
           \setlength{\topsep}{0pt}
           \setlength{\partopsep}{0pt}
           \setlength{\leftmargin}{2em}
           \setlength{\labelwidth}{1.5em}
           \setlength{\labelsep}{0.5em} } }

\newcommand{\squishend}{
   \end{list}  }
\begin{document}
\maketitle
\begin{abstract}


Software is the outcome of active and effective communication between members of an organization. This has been noted with Conway's law, which states that ``organizations design systems that mirror their own communication structure.'' However, software developers are often members of multiple organizational groups (e.g., corporate, regional,) and it is unclear how association with groups beyond one's company influence the development process. In this paper, we study social effects of country by measuring differences in software repositories associated with different countries. Using a novel dataset we obtain from GitHub, we identify key properties that differentiate software repositories based upon the country of the developers. We propose a novel approach of modeling repositories based on their sequence of development activities as a sequence embedding task and coupled with repo profile features we achieve 79.2\% accuracy on the identifying the country of a repository. Finally, we conduct a case study on repos from well-known corporations and find that country can describe the differences in development better than the company affiliation itself. These results have larger implications for software development and indicate the importance of considering the multiple groups developers are associated with when considering the formation and structure of teams.

\end{abstract}
\section{1. Introduction}
Communication, coordination, social interaction, motivation, and diversity can all affect the process and outcome of software development, influencing aspects including cost savings, development cycles, software quality, and product integration \cite{von2012carrots,liang2009software,stewart2006impact,espinosa2007team}. Communication, for example, has been suggested to shape software design \cite{conway1968committees} as ``organizations design systems that mirror their own communication structure.'' Software developed by commercial companies has been found to differ from that developed by open source communities; this may reflect the different communication structures of the two types of organizations \cite{maccormack2012exploring,syeed2013socio}. 

Similar to communication, country association may influence pertinent aspects of software development, but it has not yet been explored extensively. Country association can affect the work environment \cite{aycan2000impact} and it has been found to be more pertinent than other factors (e.g. economic development) in the adoption of organizational practices \cite{jonsson2011impact}. While broad in scope, nationalistic culture can greatly influence individual attitudes \cite{mueller2001culture} and it plays a foundational role in language use and communication. If country association influences the process and outcome of software development, this insight will have repercussions for enhancing software development and the formation of software development teams \cite{conway_small_team}.

In this work, we explore whether there are differences in the process of software development depending on country association. We ground our initial exploration of this problem by focusing on two countries: the United States of America and the People's Republic of China. As two large nation-states, both countries exhibit at the national level unique cultural dimensions. The languages spoken by the majority population from each country have different origins, with different proto-language roots. Comparing software development process differences between these two countries may indicate whether more extensive explorations of this question are warranted. In addition, the cultural model of a nation state has been suggested to be similar to that of a corporation \cite{petrovic2017comparison}. With the gap between what defines a nation state versus a corporation narrowing, we compare the differences between software development processes of companies and those of countries, exploring whether company or country contributes more to differences in development.  

We utilize GitHub to explore the process of software development; GitHub is both a social networking site for programmers and a resource for managing and sharing software, where software development activities such as updating code, tracking bugs, and reviewing code are logged. Using a curated dataset obtained from GitHub, we pose the following two research questions: 

\squishlist
    \item[] \textbf{RQ1.} Does software development on GitHub differ between repositories associated with the USA and China? 
    \item[] \textbf{RQ2.} To what extent can the software development activity found on GitHub be used to predict the country associated with a repository? 
\squishend
 
\medskip

\noindent We make the following contributions as a result:


    
     
    
     

\begin{enumerate}
    
    \item We curate a dataset of GitHub repositories separated by the country association of the developers. This dataset contains the repositories alongside their full code and change history. It also contains all of the data pertaining to different software development events. (Section 3)

    \item We model the repositories by assessing differences in development activity and find that many differences are statistically significant and may be able to differentiate the country association of a repository. In addition, we propose an end-to-end model to capture the temporal patterns in the sequence of repo activities. (Section 4)
    
     \item We extensively evaluate our proposed models to predict the country of a repository and identify the factors that contribute most to the prediction task. (Section 5) 
     
    
    \item We investigate the role of \textit{company} in software development and how company compares to country by analyzing repositories belonging to the official GitHub sites of multinational companies. (Section 6)
\end{enumerate}

We will release the anonymous repositories and their associated events data in a machine readable format to encourage further research.\footnote{\url{https://github.com/anonymous-submission/Social_Effects_of_Country_in_SW_Development}}

\section{2. Related Work}
Explorations of software repositories like GitHub have focused on insights such as what leads an individual to become a long-term developer, which team attributes seem to lead to more successful projects, and the benefits of social-networking enabled by sharing development online~\cite{cosentino2017systematic,de2019identifying,celinska2018coding,batista2017collaboration,vasilescu2015perceptions,storey2016social}. Some explorations of long-term developers and team attributes have delved into the role of a developer's geographical location~\cite{rastogi2016geographical,rastogi2018relationship,prana2020including,furtado2020successful,nadri2020relationship}. Rastogi and colleagues found that when submitters and integrators are from the same geographical location there is 19\% increase in likelihood that the pull requests will get accepted, with integrators finding it easier to work with submitters from the same geographical area. Prana and colleagues explored whether gender participation in software development differed significantly depending on country, indicating that geographic factors can pose barriers to participation in software development. As more diverse software development teams differ in development practices, if we observe differences in the software development process between countries on GitHub, these differences could be related to factors like gaps in gender participation. 

Other related work has investigated benefits relating to collaboration and communication arising from the social-networking aspects of online platforms like GitHub. Evaluations of communication patterns have received more attention than questions related to geographical location~\cite{ortu2018mining,tsay2014influence,tsay2014let,marlow2013impression,yu2015wait,guzman2014sentiment,ortu2018mining,lima2014coding}. Ortu and colleagues used 650k comments from 130k issues on GitHub to explore differences between developers; they found a relationship between communication behaviors and different types of developers, with multi-commit developers being less polite and less active in comments. Tsay and colleagues investigated how developers evaluate, discuss, and address pull requests. In many of these explorations, the focus has been on communication and collaboration behavior and how it may differ depending on factors like social connections. Given the link between communication, language, and country, differences in communication and collaboration patterns may be related to underlying country affiliations. In this work, we look at similar development activities and explore how they differ by country. 



\section{3. Curating a Country Focused Dataset}

We curate a dataset of software development activities from software repositories on GitHub, the most popular platform for software development and version control. Repositories or repos are storage locations for software projects; repositories enable users to contribute to code development and collaborate with other users. The process for collecting, organizing, and refining the dataset involved several steps including evaluating repo activity, assessing repo profile details, extracting user information, and aggregating country identifiers. We describe the process in more detail below.

\medskip

\textbf{Step 1: Curating by Event.} 
GitHub taxonomizes the code development process through events or repository activities (we use \textit{event} and \textit{activity} interchangeably throughout this paper). All repository-level events are openly available for public repositories. We collect these events in the form of a quadruplet, $(a,e,r,t)$, i.e., actor $a$ takes action of type $e$ on repository $r$ at timestamp $t$. The types of events or activities included in our study are described in Table \ref{tab:events}. We collect all event data over a span of 3.5 years, from January 1, 2017 to June 30, 2020 (inclusive). As software changes quickly with major and minor releases capable of occurring within a single year, we hope this snapshot in time of the past 3.5 years will give us a view of differences that are reflective of recent activities. 
\medskip

\textbf{Step 2: Curating by Repository.} After this initial collection, we further refine the set of repositories in two ways. First, we would like to ensure that we are looking at repositories holistically, meaning that we have the entire life or history of the repositories we are exploring as much as possible. Therefore, we consider only repositories which contain a \textit{Create} event. The presence of a Create event indicates the repository was initiated within the collection period and that we have the entire life of the repo up until June 30, 2020. Second, we keep only repositories with $50$ or more events. This number ensures an appropriate amount of activity for insight into interaction patterns. We obtain 7.4k repos. 



\begin{table}[t!]
\setlength\tabcolsep{4pt}
\caption{\label{tbl_git_events_description} GitHub events included in our study (14 event types). These describe a project's development process.}
\vspace{-3mm}
\begin{tabular}{l p{5cm}}
\textbf{Event Type}& \textbf{Description} \\
\toprule
\textbf{Create} & Triggered upon creation of a branch or tag. 
We use this to know the genesis of a repository. \\
\midrule
\textbf{CommitComment} & A comment is made for a commit. \\
\midrule
\textbf{Push} & One or more commits are submitted to a repository. \\
\midrule
\textbf{Watch} & Triggered upon any user (contributor or not) starring the repository. \\
\midrule
\textbf{Fork} & The repository is forked by a user. \\
\midrule
\textbf{IssueComment} & A user reports an issue, often a bug report or enhancement request. \\
\midrule
\textbf{Issues} & Triggered when an issue is modified, being opened, assigned to a user, or closed. \\
\midrule
\textbf{PullRequest} & This event triggers when a user submits a pull request, which outlines the changes made to a branch.  \\
\midrule
\textbf{ReviewComment} & Any action pertaining to the comments of a pull request occurs. \\
\midrule
\textbf{Delete} & A branch or tag is deleted. \\
\midrule
\textbf{Gollum} & A repo's Wiki page is edited. \\
\midrule
\textbf{Member} & Whenever members are added or removed from a repository. \\
\midrule
\textbf{Release} & A release occurs. Releases signify project milestones, often accompanied by executables.  \\
\midrule
\textbf{Public} & The permissions change on a repository such that it becomes public. \\
\midrule
\bottomrule
\end{tabular}
\label{tab:events}
\end{table}


\medskip

\textbf{Step 3: Curating by User.} Finally, we utilize user profile details to identify the country and companies associated with each repository. Collecting user profiles via the GitHub API, we identify attributes such as \textit{location and company} for each user who engages with one of the repositories of interest. The location field in a user's profile is a free text input, meaning it is noisy and inconsistent e.g., it varies from city, university to coordinates. To address this issue, we use \textit{Nominatim}~\cite{haklay2008openstreetmap} to convert the location string into latitude and longitude and calculate the country of the user from there. We assign each repo to a country if over 50\% of its contributors come from the same country. We then extract those repositories which are associated with either the USA (N=1,451) and China (N=1,125) for a total of 2,576 repositories.







\subsection{3.1.  Initial Analysis of Repository Differences}\label{Exploring_Repo_Activities}
In this section, we examine the high level differences in activity patterns between country groups. Prior to any in-depth analyses, we observe that \textit{Watch} events are highly frequent compared to rest of the event types and we can relate this to the fact that any user on GitHub can star a repo, even in passing. Due to the low cost of entry to perform this event, we do not include it in the analyses following this paper unless stated otherwise. 
Below we describe differences observed across the combined repository activity, the most active repositories from each country, and in the topics described by the repositories from each country. 

\begin{figure}[t!]
    \centering
    \subfigure[With Watch event.]{
    \hspace*{-1cm} 
        \includegraphics[width=0.24\textwidth]{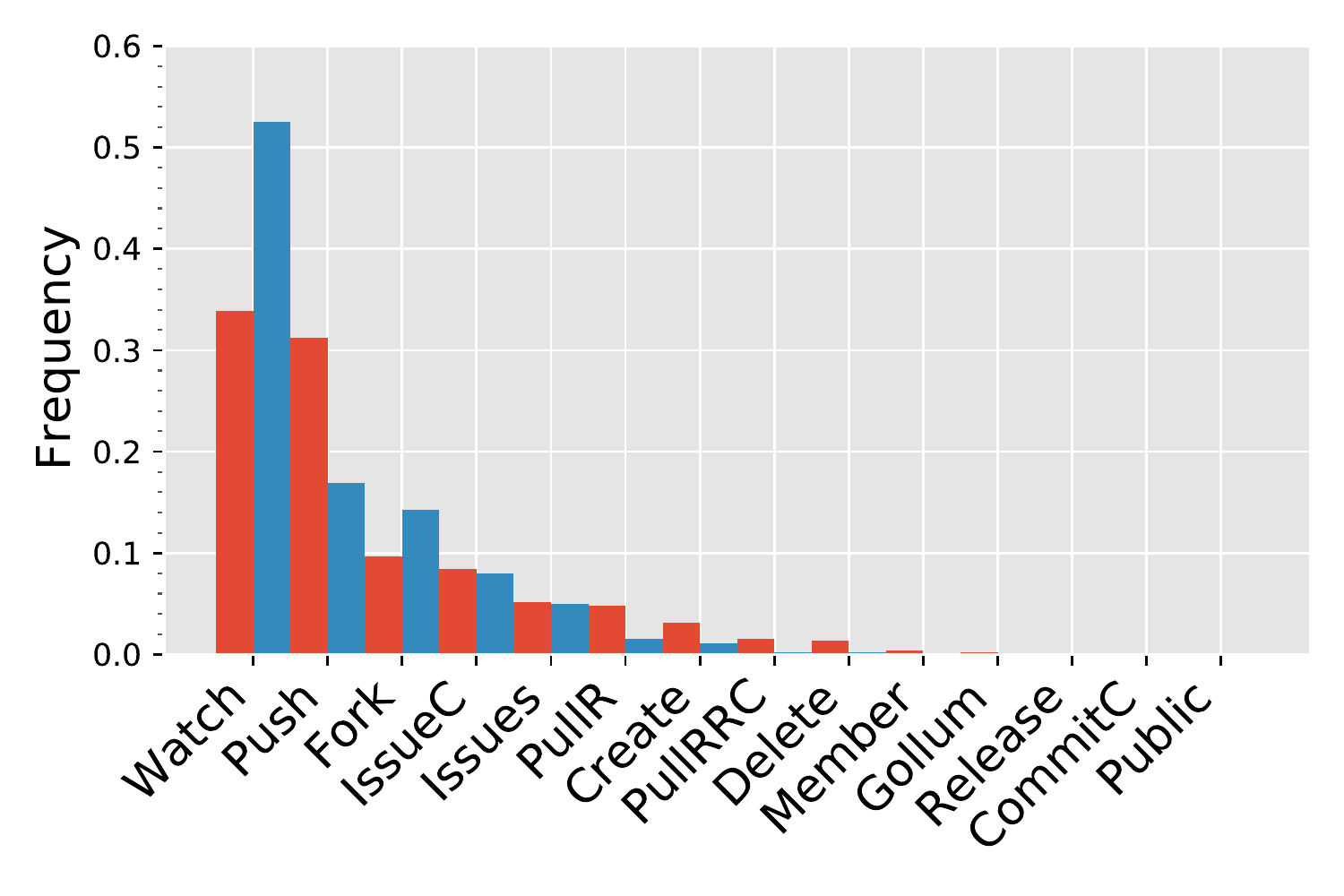}
    }
    \subfigure[Without Watch event.]{
        \includegraphics[width=0.24\textwidth]{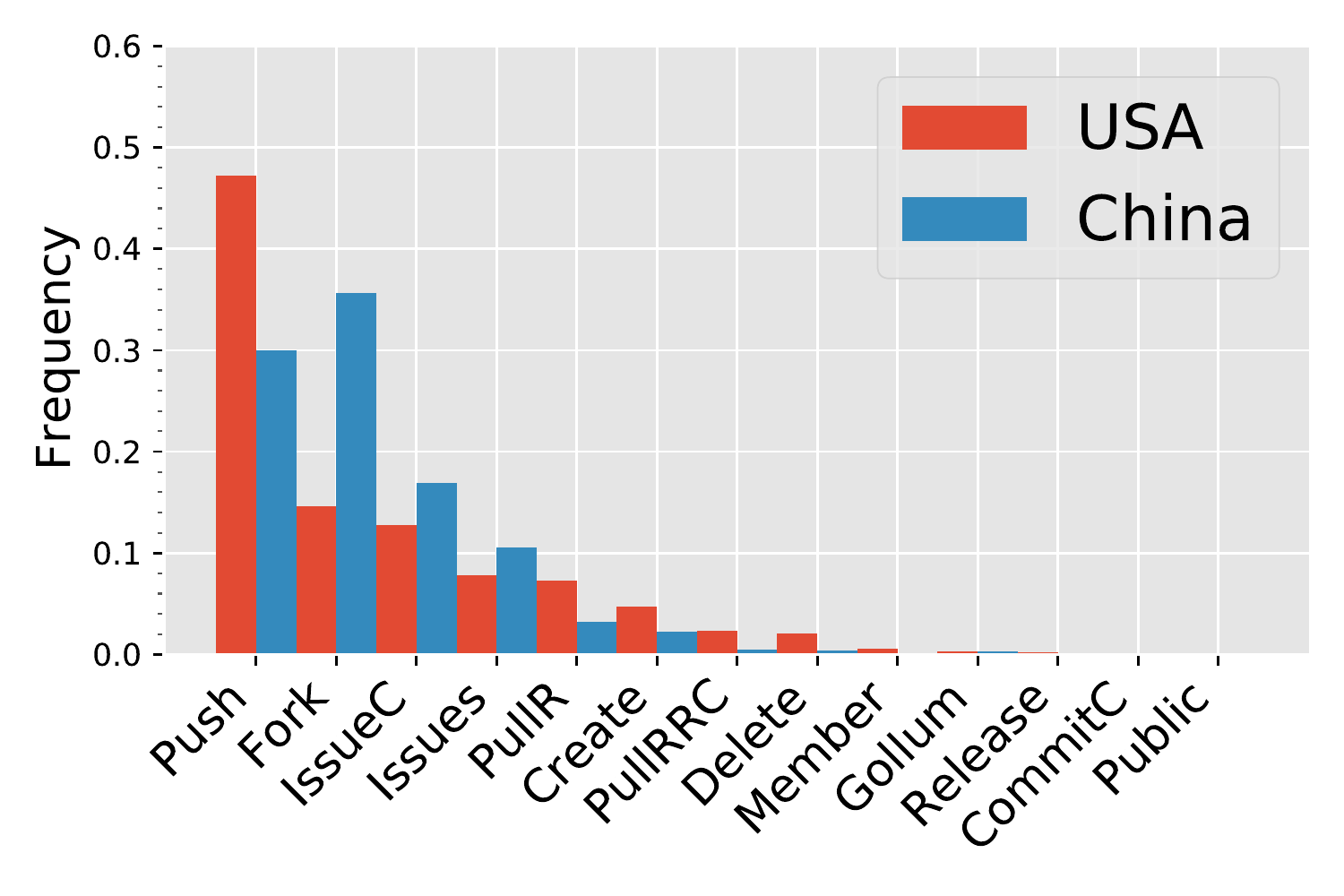}
    }
    \setlength{\belowcaptionskip}{-11pt}
    \caption{Combined distribution of events in Repositories. }
    \label{event_dist_us_china}
\end{figure}

\medskip
\noindent\textbf{Differences in Combined Repository Activity.} First, we look at all repository activity combined and explore how it differs between countries. The results of this analysis are presented in Figure \ref{event_dist_us_china}, which shows the event type distribution for repos across the USA and China. For informational purposes, we include the distributions both in the presence and absence of the Watch event. We observe that the Watch and Fork events are common in repos originating from China while the Push event is the dominant activity in USA repos. To quantify this difference, we calculate KL divergence scores that are 0.16 and 0.2 in the presence and absence of the Watch event respectively. This indicates a significant difference between the two probability distributions, where the KL divergence score of two identical distributions is zero.

\begin{figure}[tb!]
    \centering
    \subfigure[In US repos, IssueComment is more visible.]{
        \includegraphics[width=0.22\textwidth]{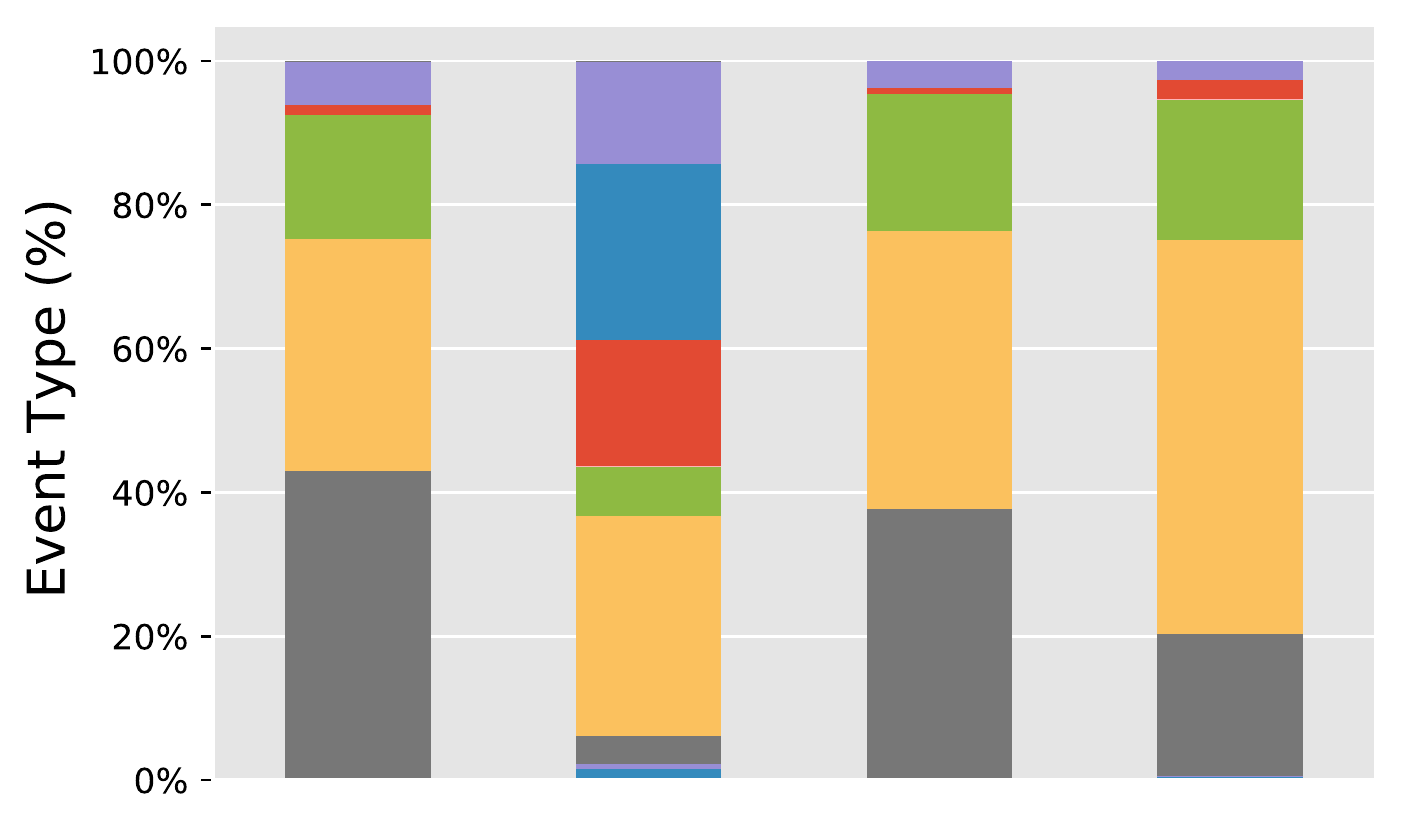}
    }
    \subfigure[In China repos, ForkEvent is dominant.]{
        \includegraphics[width=0.22\textwidth]{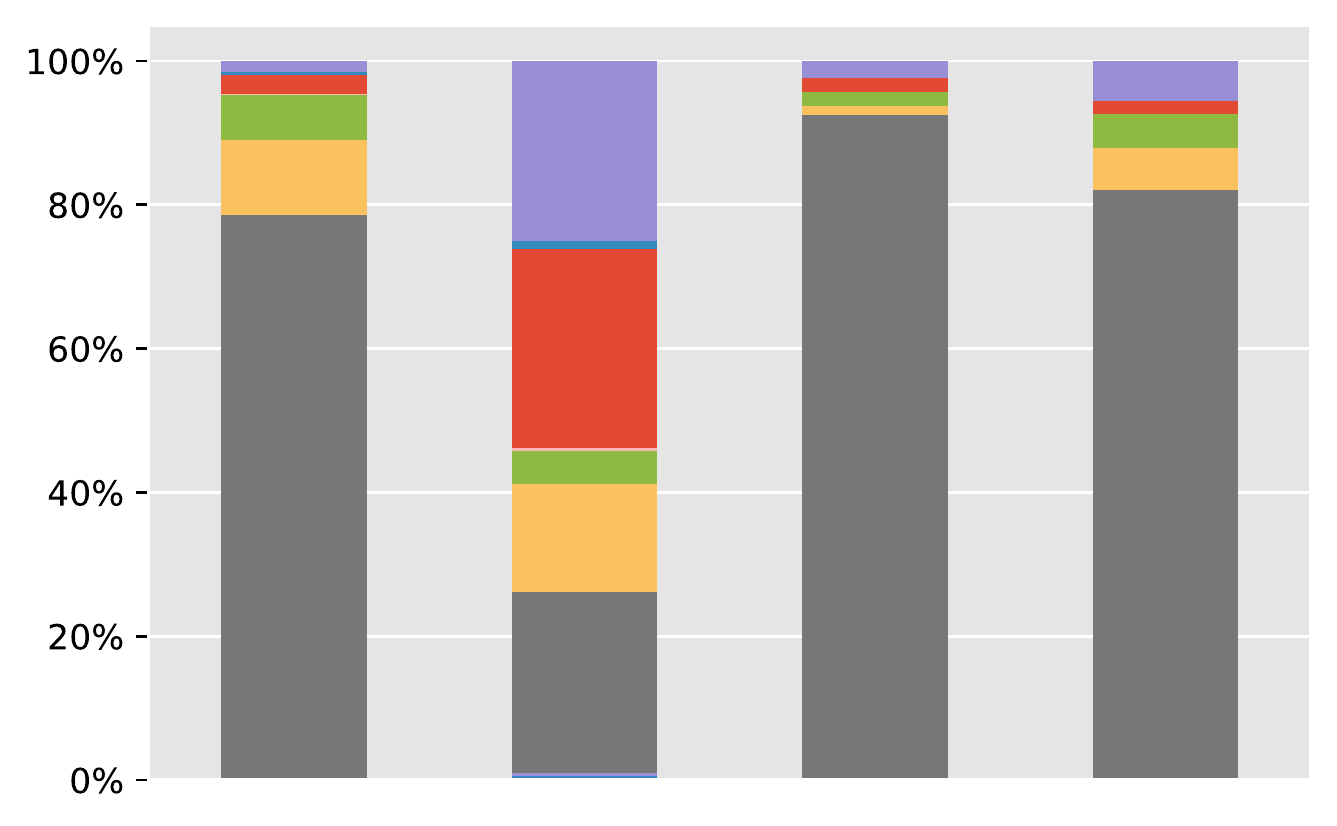}
    }
    
    \includegraphics[width=0.4\textwidth]{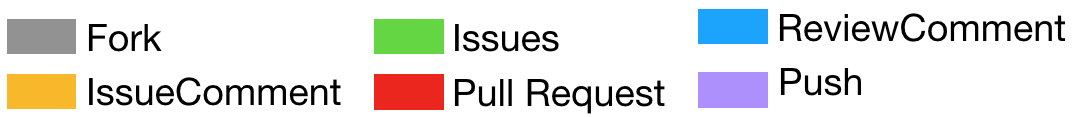}
    \setlength{\belowcaptionskip}{-11pt}
    \caption{Overview of active repositories by event fingerprint (repos w/ most number of events.) }
    \label{fig_top_repos_events}
\end{figure}

\medskip
\noindent\textbf{Analysis of Active Repositories.} Next, we visualize the distribution of events among the four most active repositories in China and the USA in Figure~\ref{fig_top_repos_events}. This reveals a stark difference in the activity patterns of these top repositories across these two different groups. First, \textbf{ForkEvent}s (\emph{grey}) are predominant in the repositories associated with China while \textbf{IssueCommentEvent}s (\emph{yellow}) are favored by US repositories. This illustrates that there are distinctions in the way developers who have different associations or memberships interact with software repositories on GitHub. We may be able to use these distinctions to indicate the country of a repository and its developers.

\medskip
\noindent\textbf{Differences in Repository Topic Distribution.}
Finally, we look at differences in the topic distributions of repositories associated with different countries. To do this, we extract repo descriptions from \textit{Create} event payload. We notice that some of the repos (42 out of 1,125) associated with China contain non-English words where we convert the repos' description to English using langid and translator.translate Python libraries to detect Chinese characters and translate them into English resp. We also find that topics like ``neural" and ``network" are more frequent in repos associated with USA while ``implementation", ``tensorflow" and ``PyTorch" are favored by Chinese repositories. In the next section, we show that the topic distribution computed by the LDA method is significantly different among these two groups. This suggests distinctions in the topics of interest for repositories associated with China versus repositories associated with the USA.


\medskip
\noindent\textbf{Summary.} Through our high level study, we find that there exist differences in code development activities in aggregate and separated by individual repos. We also show that developers with different background may favor different topics. This motivates us to conduct a fine-grained study of development activities in order to propose a predictive model. 


\section{4. Modeling Development Activity}\label{prediction}

\begin{figure*}[t!]
    \centering
    \subfigure[Stars\textbf{**}]{\includegraphics[width=0.24\linewidth]{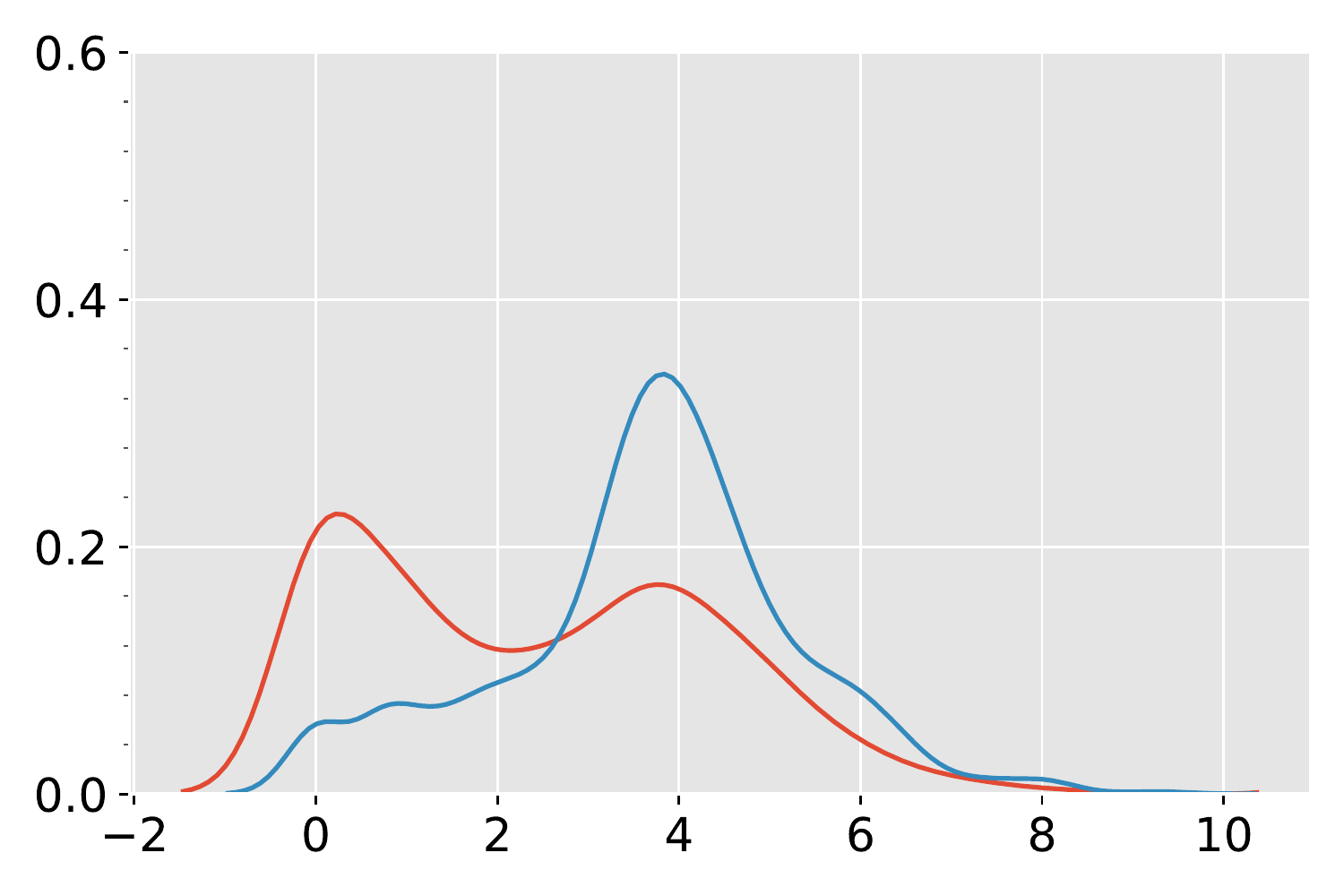}}
    \subfigure[Forks\textbf{**}]{\includegraphics[width=0.23\linewidth]{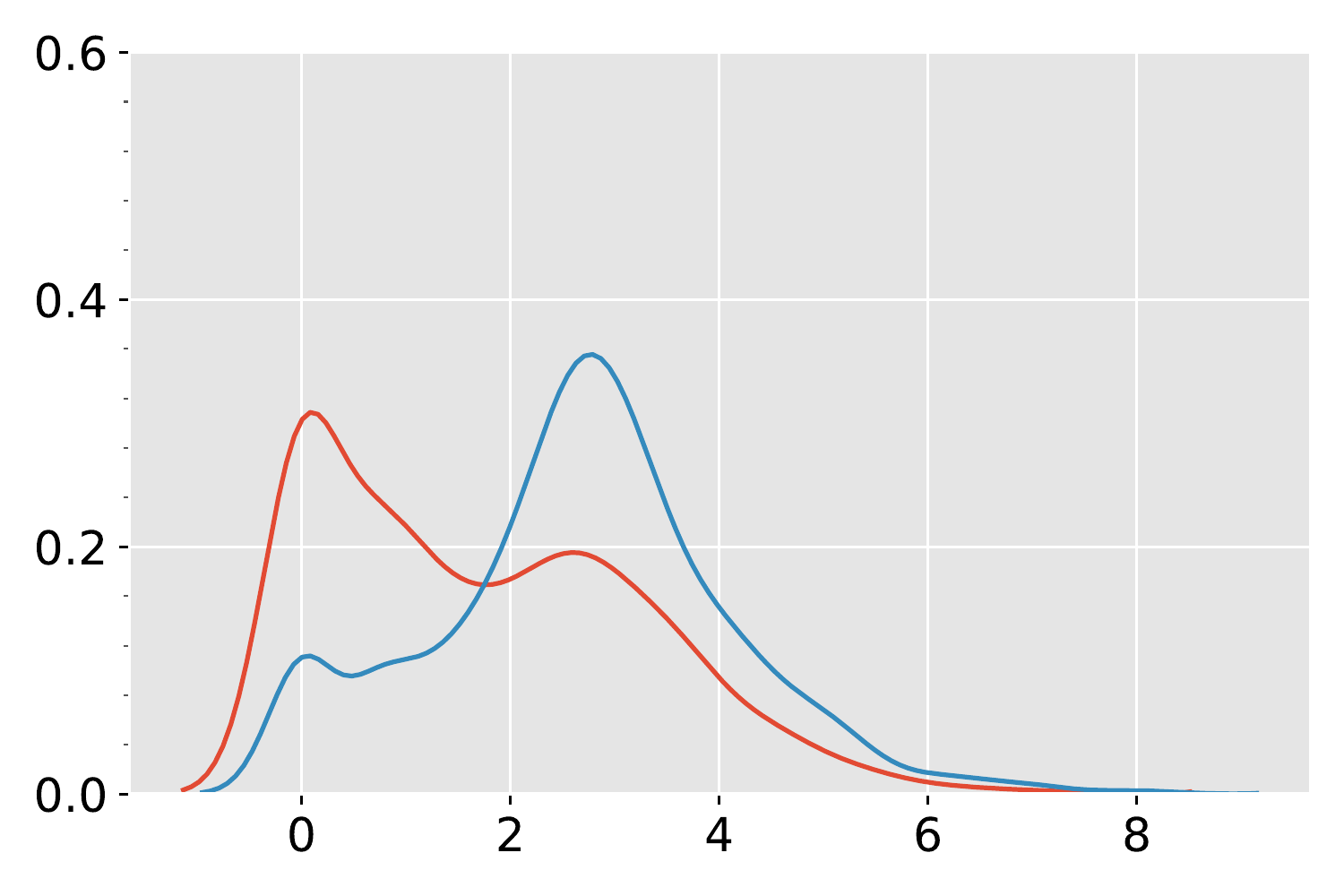}}
    \subfigure[Open Issues\textbf{*}]{\includegraphics[width=0.23\linewidth]{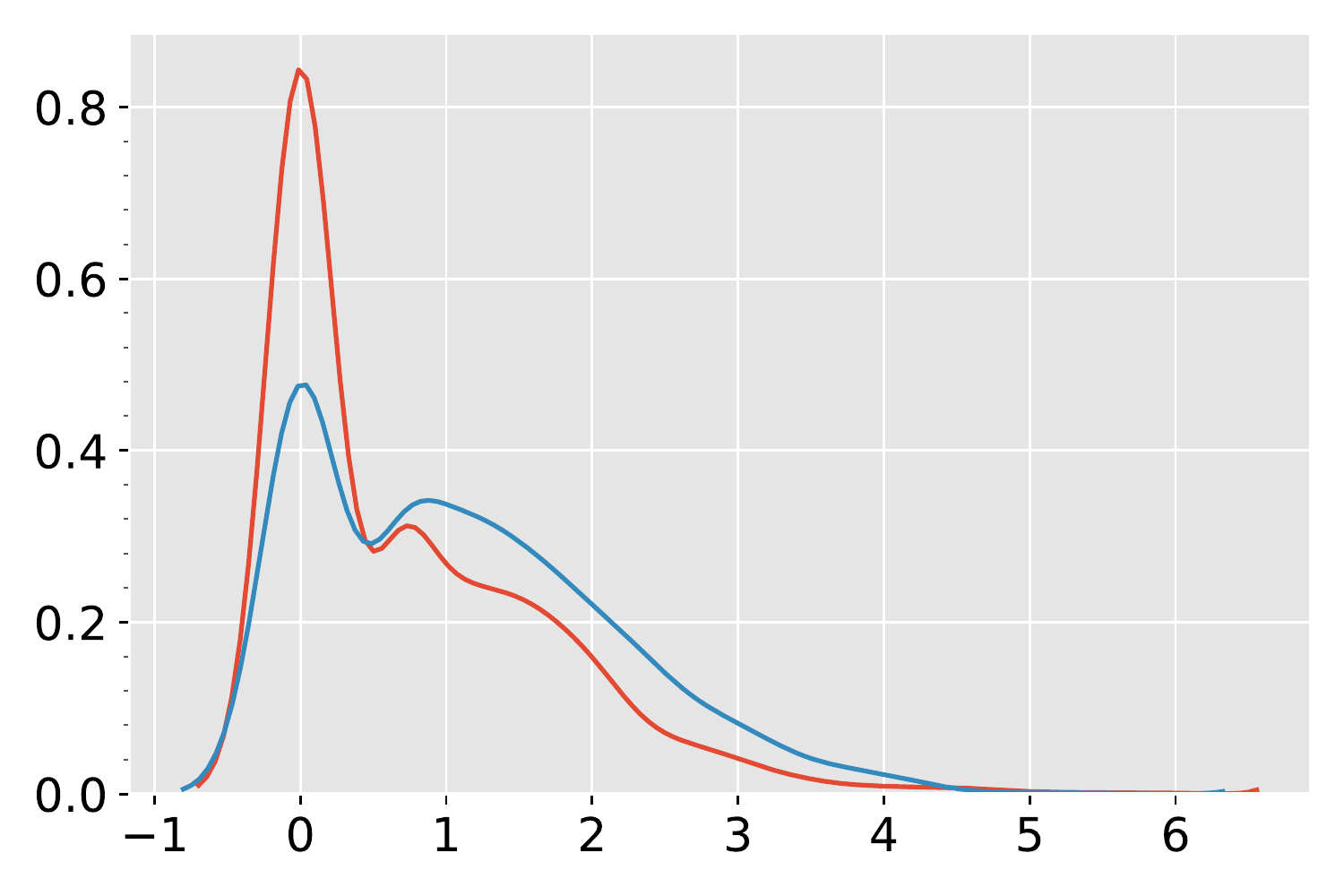}}
    \subfigure[Comment Length\textbf{**}]{\includegraphics[width=0.23\linewidth]{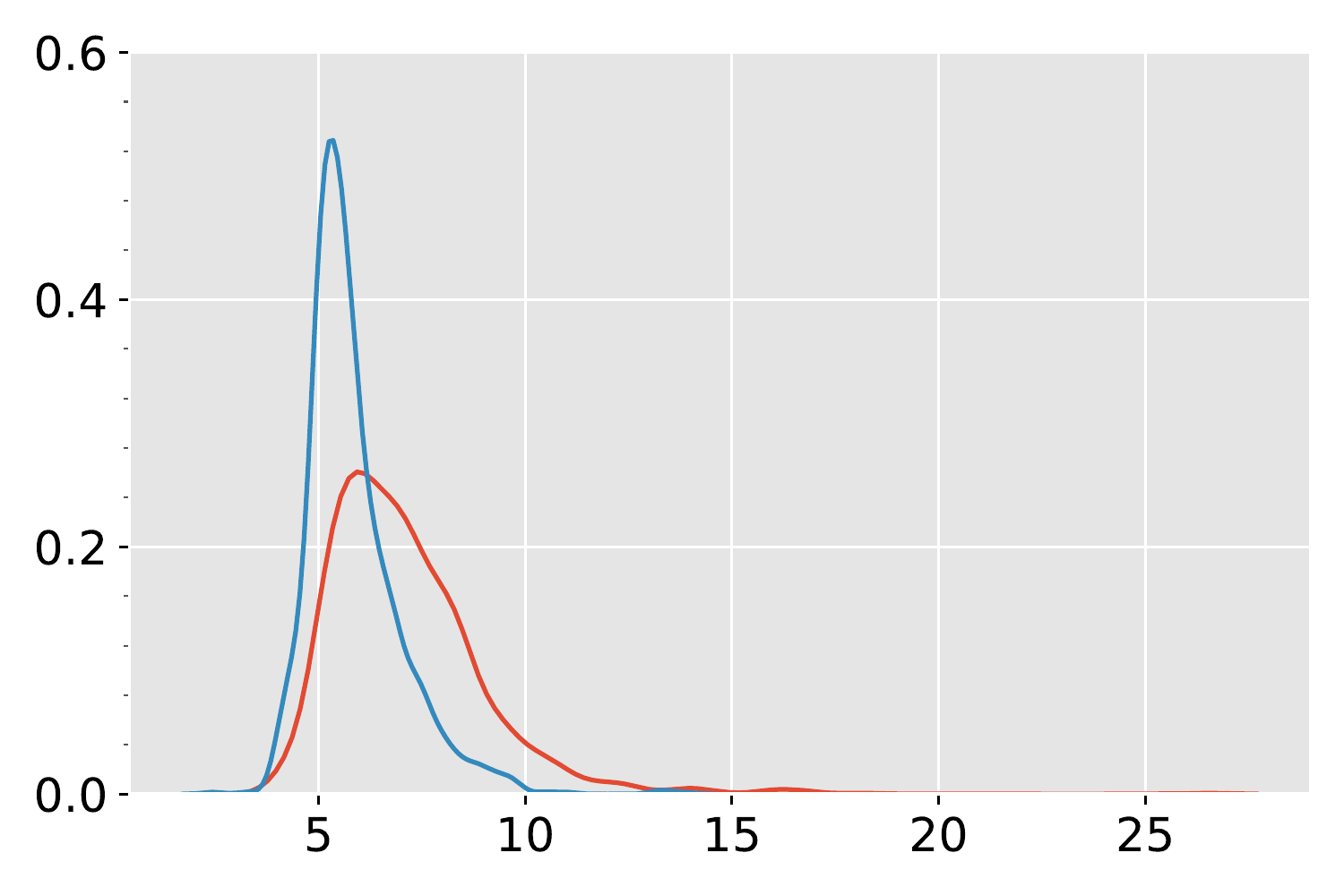}}
    \quad
    \subfigure[Event Interarrival Time\textbf{**}]{\includegraphics[width=0.23\linewidth]{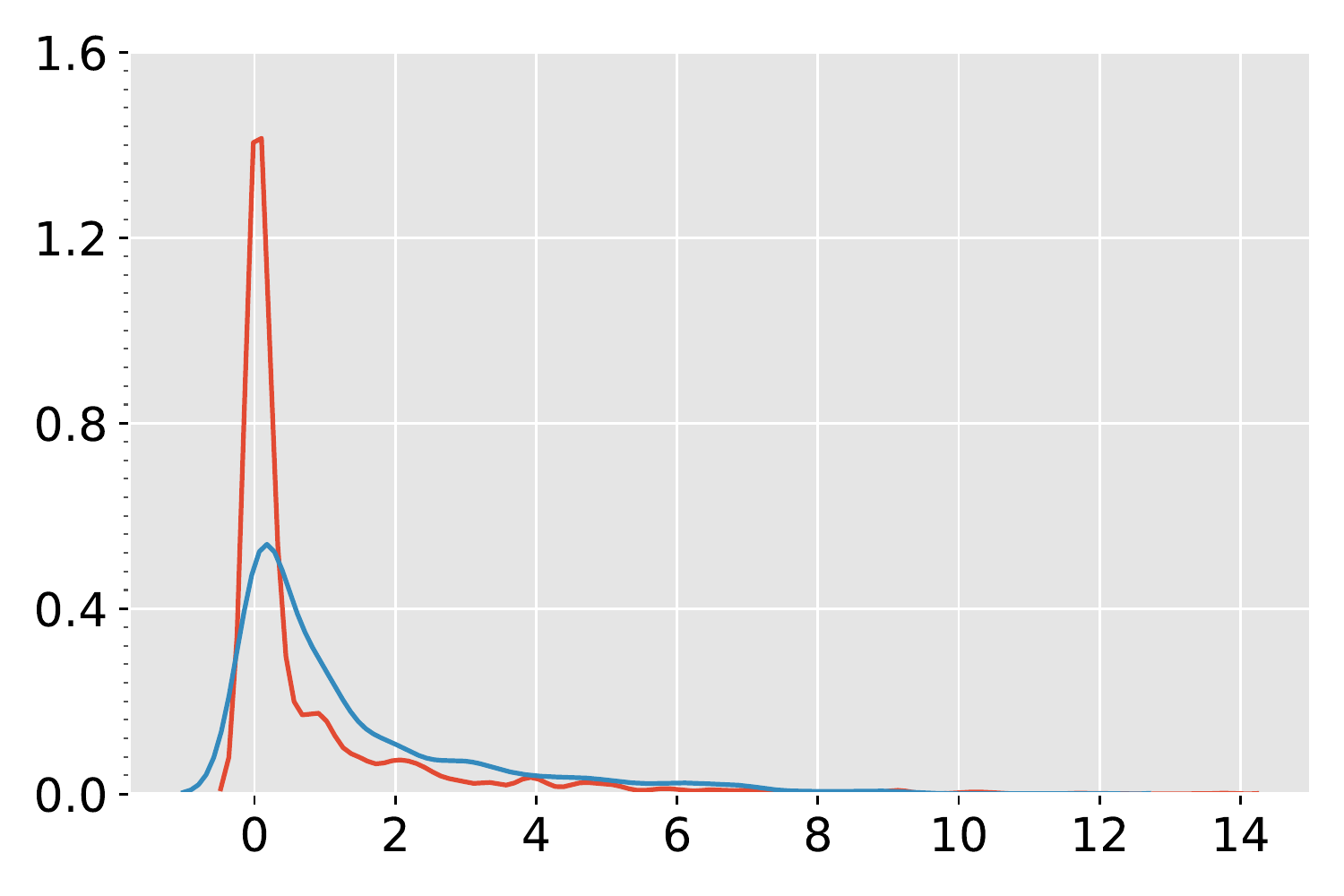}}
    \quad
    \subfigure[Leaders\textbf{*}]{\includegraphics[width=0.23\linewidth]{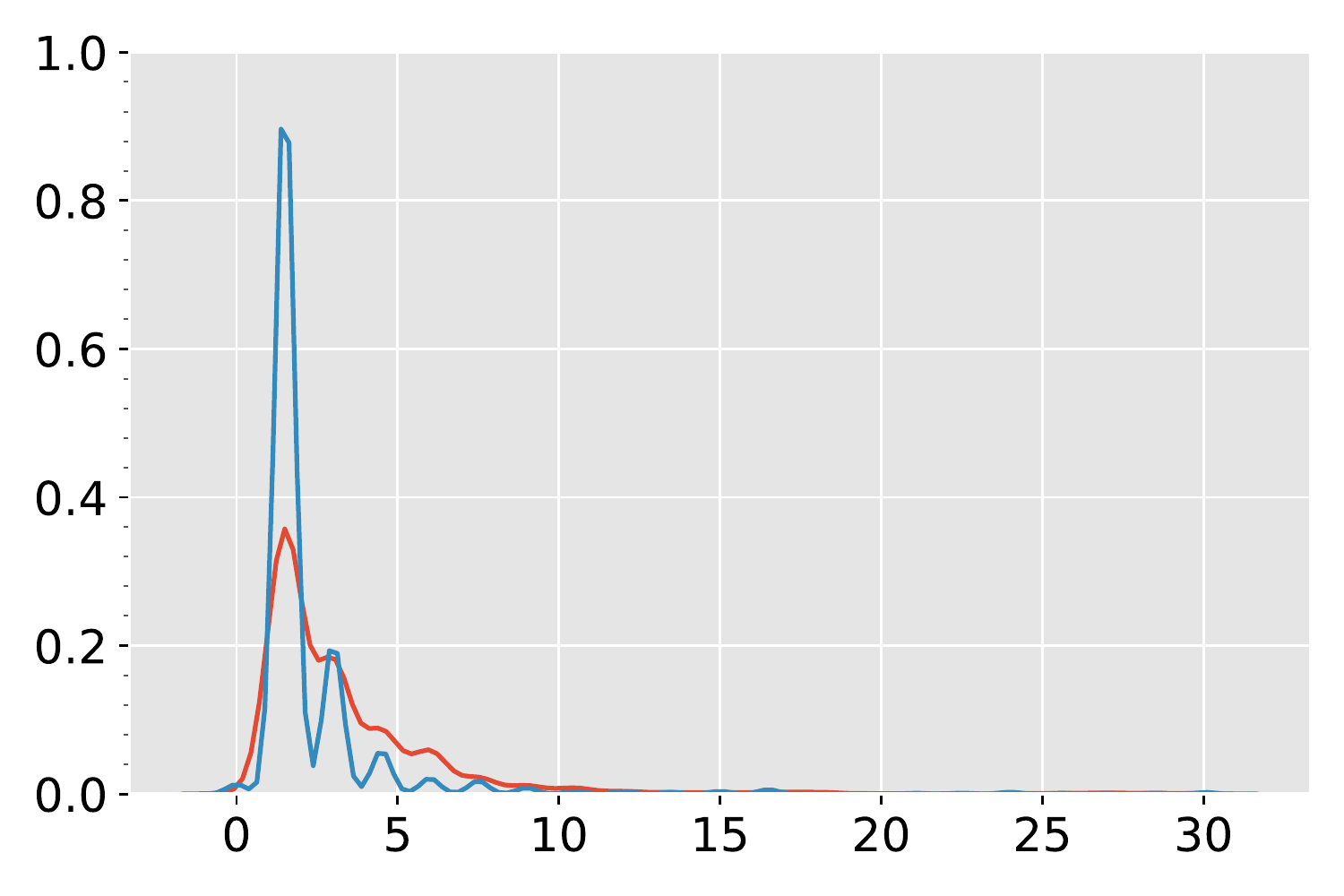}}
    \quad
    \subfigure[Watcher/Contributor Jaccard\textbf{**}]{\includegraphics[width=0.23\linewidth]{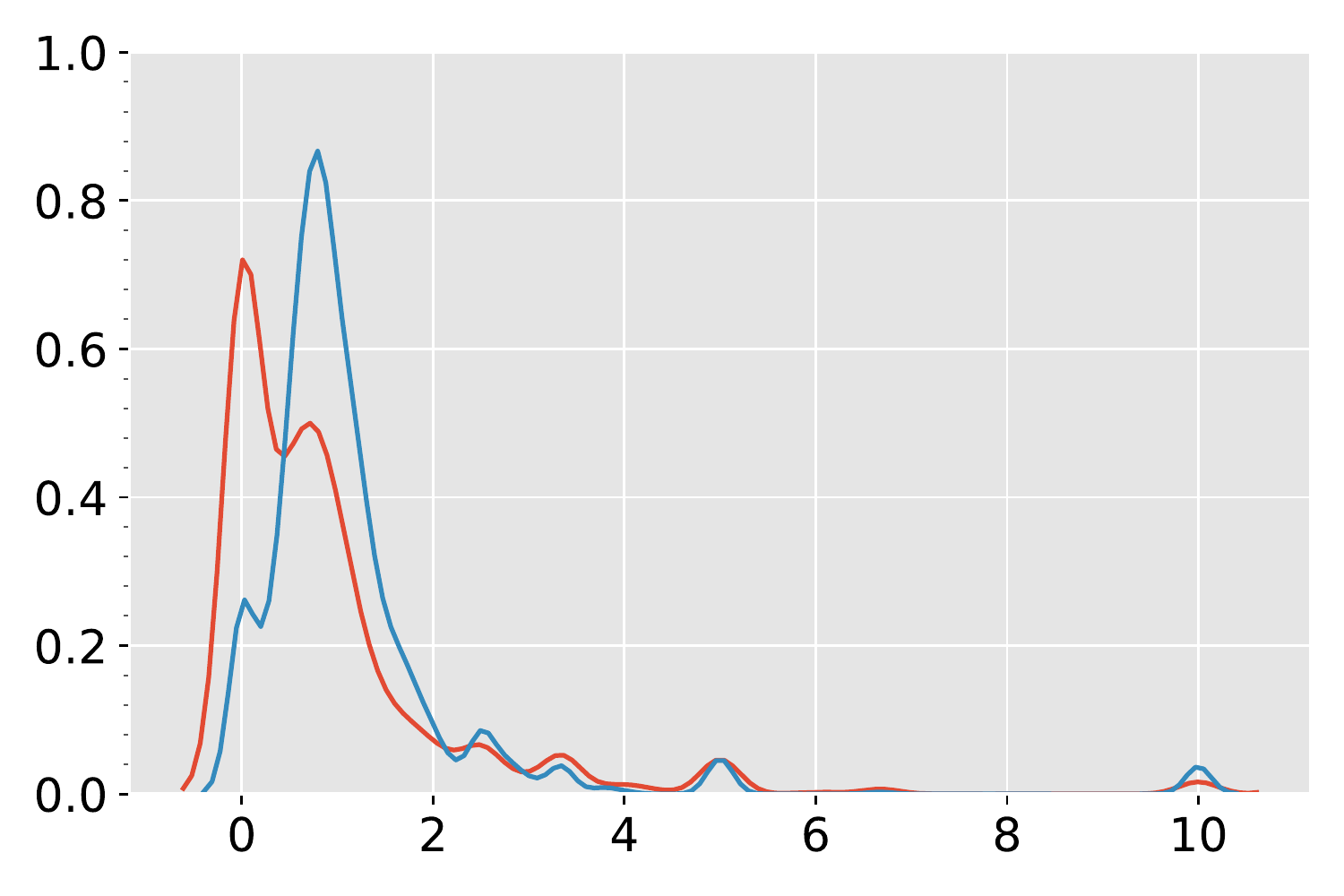}}
    \quad
    \subfigure[Topic 1 Affinity\textbf{**}]{\includegraphics[width=0.23\linewidth]{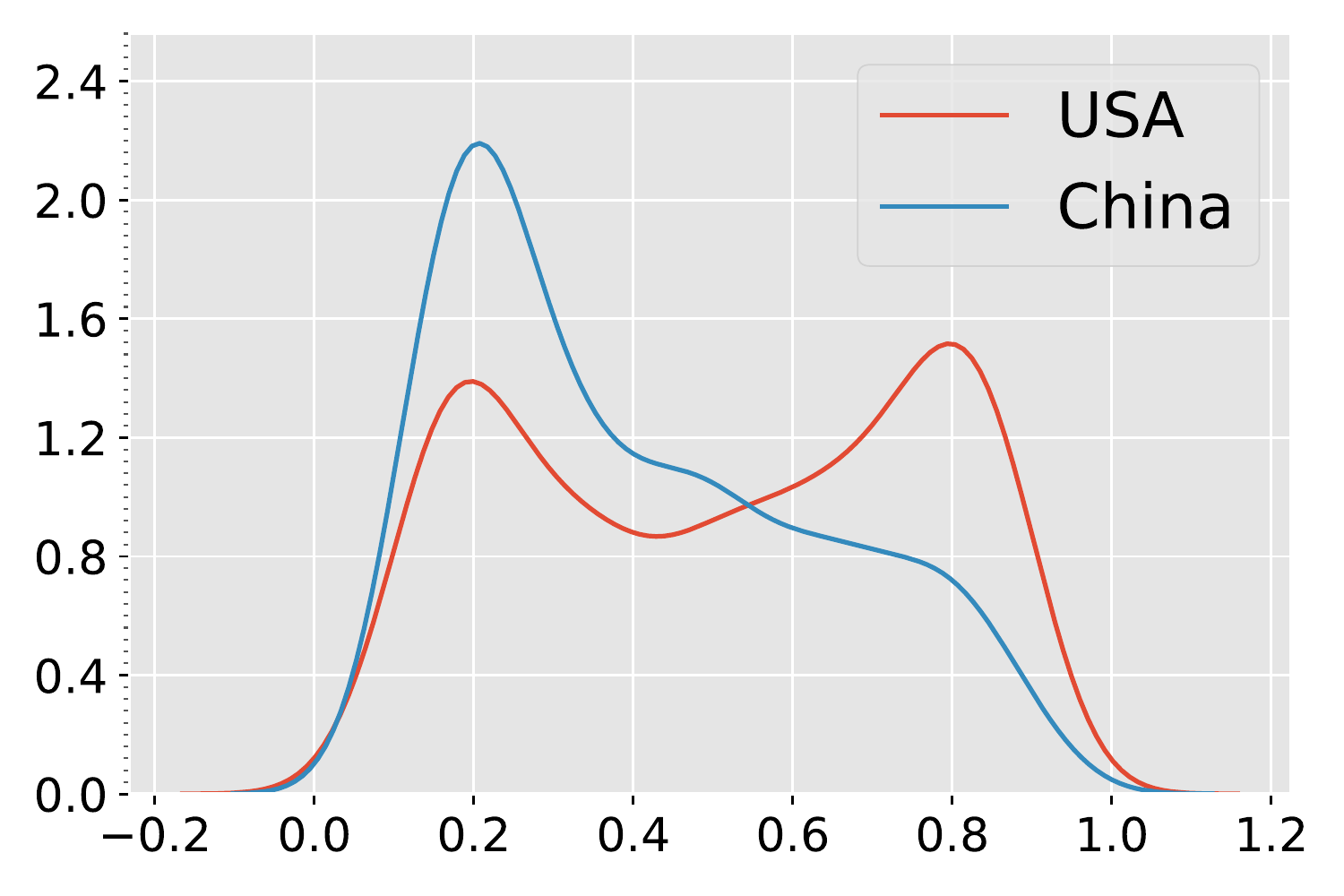}}

    \setlength{\belowcaptionskip}{-11pt}
    \caption{Distribution of repo profile features across repositories in the US and China. A $z$-test was performed to assess the difference in the distributions. A \textbf{*} indicates that we reject $H_0$--that the distributions are the same--at $\alpha=0.005$, and \textbf{**} at $\alpha=10^{-4}$. Jaccard's x-axis was scaled by a factor of 10, only in the visualization.}
    \label{fig_feature_distribution}
\end{figure*}

To explore the effects of country on software development, we first develop a set of features we can use to model the development activity of repositories. Using the results from the prior section, we explore several representations of repositories based on \textit{repository profiles}, frequency of different \textit{types of events}, and \textit{sequences} of repository activities; the features from these three sets of representations are concatenated as the feature vector for the model explored in Section 5.  


\medskip

\noindent \textbf{Repo Profile Features.} 
We compute features that are closely related to the profile or overall representation of a repository including its users and activities. The attributes we utilize are described below. To motivate the suitability of these features, we perform a statistical test to see if the values between USA and Chinese distributions are significantly different. Here, we perform a $z$-test~\cite{casella2002statistical}, which tests $H_0:$ the means of the two distributions are the same. The results and visualizations of the distributions are given in Figure~\ref{fig_feature_distribution}. 

\noindent \textit{Star count.} This is computed as the number of stars a repository has been given by GitHub users. While both countries exhibit a somewhat bimodal distribution, USA repos have, on average, fewer stars than Chinese repos.

\noindent \textit{Fork count.} This is the number of forks a repository has received. Similar to \textit{star count}, both countries exhibit a somewhat bimodal distribution; however, USA repos have, on average, fewer forks than Chinese repos.

\noindent \textit{Open issues count.} This is the number of issues that have been opened in the lifetime of a repository. Most repos from both countries have zero open issues.

\noindent \textit{Length of Comments in Commit events.} Commits are paired with comments to describe the changes made to the code base. We take the average length of all commit comments made on a repository. Comments that were detected as being in Chinese were first translated to English for a fair comparison. Developers associated with Chinese repos appear to write shorter comments.  

\noindent \textit{Event Inter-arrival Time.} This is computed as the median of the time between events in the repository, in days. Given repo $r=\{e_1,...,e_n\}$, $iat_r=Median\{\left(t_{e_{i+1}}-t_{e_i}\right) \forall i \in [1,...,n]\}$ where $t_{e_i}$ is the occurrence \textit{time} of event $e_i$. We take the median rather than the average given that it is common for more development events to take place in the early life of a repo on GitHub. USA repos tend to have a much shorter inter-arrival time between events.

\noindent \textit{Number of Leaders.} The number of unique users with Push and Pull request access. The curve is jagged in both countries; China has more repositories with just one leader.

\noindent \textit{Jaccard similarity of actors.} This feature shows the Jaccard similarity between watchers and contributors where a low Jaccard may be a sign of higher outside interest due to lower overlap between watchers and contributors. China has a greater Jaccard similarity between its contributors and leaders. This suggests there may be stronger outside interest in the repositories associated with the USA. 

\noindent \textit{Topic Distribution}. This calculates the distribution of topics from repos description using LDA technique. Give k number of topics, it produces a probability distribution over k topics for each repo. This suggests that repos from different organization may favor different topics. We set k=2 with this intuition that each topic represent repos from either country. 

\medskip
\noindent \textbf{Repo Activity Features} Secondly, we compute features that are based on the specific activity for a repository. As demonstrated in  Figure~\ref{fig_top_repos_events}, repos from different countries may have different activity fingerprints. We determine a repository's activity fingerprint with each activity as a percentage or frequency of overall activities, calculated as $\{(|e_j|\big/n) \forall j \in types\}$. As a reminder, the full list of activity types can be found in Table~\ref{tab:events}.

We evaluate differences in activity fingerprints by performing a $z$-test on the percentages, assessing the level of significance between the USA and Chinese distributions. We find that almost all activity features are statistically significant at the $\alpha=10^{-4}$ level. The exceptions are 
``ReleaseEvent'', ``GollumEvent'', ``CommitCommentEvent'', and ``PublicEvent''. We include all the activity features in our model for the sake of generality and investigate their effectiveness through feature ablation tests.

\medskip
\noindent \textbf{Sequence Embedding Features.}
Finally, we explore the potential of sequences of activity. The features we described above which were extracted from repository profiles and based on the frequency of repository activities do not capture the temporal connections which exist between events. For example, the two sequences of  $r_1=\{$IssueComment, IssueComment, Issues, Issues, Push, Push$\}$ and $r_2=\{$IssueComment, Issues, Push, IssueComment, Issues, Push$\}$ appear to have identical events probability distributions. However, the order of the events can indicate vital differences in software development practices. In the sequence for $r_1$, contributors appear to collect issues first and then respond to several issues at once while $r_2$ contributors address each reported issue right after its occurrence. The potential insight these subtle behaviors can provide leads us to propose a sequence embedding approach to capture this nuance in an end-to-end manner. 

Since Recurrent Neural Networks (RNNs) are capable of modeling order dependency in sequential data, we use RNNs (LSTMs cell) as the neural component of the proposed sequence embedding. Inspired by recent advances in Variational \textit{Recurrent} Auto-Encoders (VRAE) \cite{fabius2014variational}, we map sequences of activities associated with each repo to a latent vector representation in an unsupervised fashion. Note that Variational Auto-Encoders can also serve as generative models by producing probability distributions over each of the latent vector variables and enabling sampling from these distributions to generate new samples. However, our main intuition here is to transform the sequence of activities into an embedding space so that sequences with similar patterns are closer in the new space and sequences with less similar patterns are distanced. In the following, we explain the steps in sequence embedding task.


\noindent \textit{1. Event Sequence Learning.} Without loss of generality, we consider repo $r_i=\{e_1,e_2,...,e_n\}$ where the encoder transforms $r_i$ into latent vector $z_{r_i}$ and the decoder reconstructs $r'_i$ from  $z_{r_i}$ so that $r'_i$ replicates $r_i$ as close as possible. To put this concept into a learning process, the optimizer aims to maximize the likelihood of observing $z_{r_i}$ given $r_i$ i.e. $p(z_{r_i}|r_i)=\frac{p(r_i|z_{r_i})}{p(r_i)}$ while the likelihood $p(r_i)= \int p(z_{r_i}) p(r_i|z_{r_i})dz_{r_i}$ is intractable given that the latent vector $z_{r_i}$ size is $k$ and it is hard to evaluate multiple number of integrals as the size of $k$ increases. \cite{rezende2014stochastic} proposes to approximate $p (z_{r_i}|r_i)$ by Gaussian distribution of $q (z_{r_i}|r_i)$. 



As a result, the optimization problem is reduced to minimize the following loss function: 
\begin{equation}
\mathcal{L} (\theta; r_i)= -D_{KL} (q(z_{r_i}|r_i), p(z_{r_i})) + \mathbb{E}_{q(z_{r_i}|r_i)} [\log p_{\theta}(r_i|z_{r_i})] 
\end{equation}

This is composed of two components: 1) the KL-divergence between the distribution learned in latent space ($p(z_{r_i})$) and the Gaussian distribution $\mathcal{N}$($q (z_{r_i}|r_i)$); 2) the reconstruction error which is based on the main intuition behind auto-encoders models that the input sequence and output vector must be similar. So, the MSE loss between $r_i$ and $r'_i$ (input and decoded vectors resp) is calculated. 

\noindent \textit{2. Recurrent Neural Network.} The overall architecture of sequence embedding is demonstrated in Figure \ref{fig_sequence_emb_archi}. The \textit{Encoder} contains $n$ (sequence length) number of recurrent connections. At each time step ($t$), the hidden state ($h_t$) is calculated based on previous hidden state ($h_{t-1}$) and the current event data ($e_t$). The last hidden state ($h_n$) is passed to the encoder-to-latent layer. This layer performs a linear operation over $h_n$ to obtain the mean and standard deviation of the distribution from which the latent vector $z$ is sampled:

\begin{figure}[tb!]
    \centering
    \includegraphics[width=0.6\linewidth]{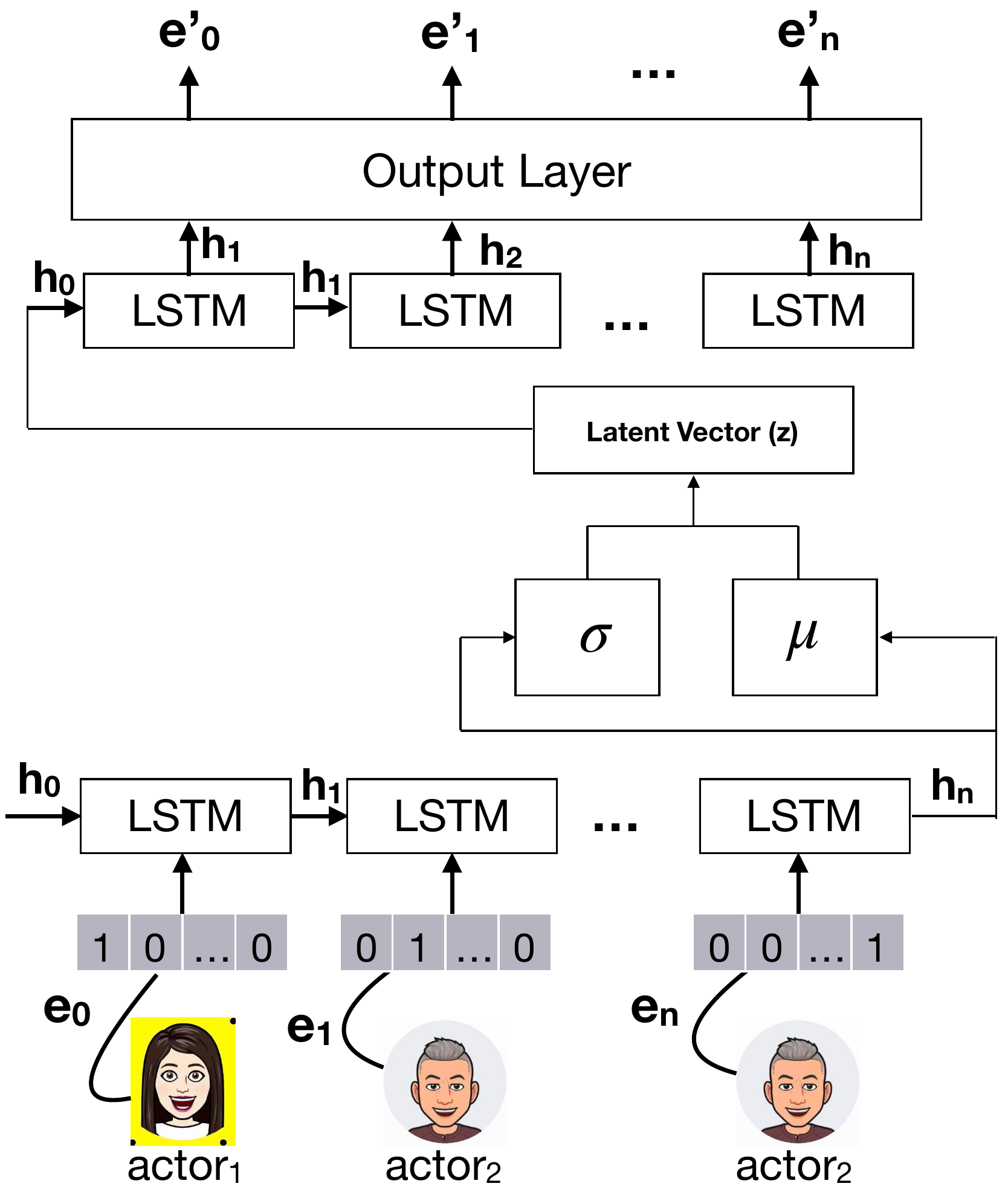}
    \caption{Overall architecture of the sequence embedding.}
    \label{fig_sequence_emb_archi}
    \vspace{-2em}
\end{figure}

\begin{align*}
    h_t=LSTM(h_{t-1}, e_t)\\
    \mu_{z}=W^T_{\mu}h_n + b_{\mu}\\
    \log(\sigma_z)=W^T_{\sigma} h_n + b_{\sigma}
\end{align*}

Hence, $z$ is sampled as $z=\mu+\sigma \epsilon$ with $\epsilon\ \mathord\sim \mathcal{N}(0,1)$. On the \textit{Decoder} side, the first hidden state is initialized by a linear operation over $z$ and the rest are updated similar to classic RNNs such that the decoder inputs ($x_t$) are initialized to zero. The output layer is a linear operation that maps decoder outputs (i.e., $h_t$) to the sequence to reconstruct the input events ($e_t$). 

\begin{align*}
    h_0=tanh(W^T_z z + b_z)\\
    h_t=LSTM(h_{t-1}, x_t)\\
    e_t=sigm(W^T_{out}h_t+b_{out})
\end{align*}

\noindent $W$ and $b$ are weight and bias; learnable parameters.

\noindent \textit{3. Input Layer}. Given there are 14 types of GitHub events, we map each event into a one-hot vector which is further fed into recurrent network. Therefore, the number of input features for RNN is 14 where the number of time steps depends on length of the sequence (number of events).

\section{5. Predicting the Country of a Repository}\label{sec_rq2_results}
In this section, we aim to evaluate whether the repo representation we developed in previous section can differentiate between repositories associated with China and US. Before turning to experimental results, we introduce a notion of repo clustering that is the result of power-law distribution in code development activities.

\begin{figure}[tb!]
    \centering
    \subfigure[Contributors per repo]{
      \hspace*{-1cm} 
        \includegraphics[width=0.24\textwidth]{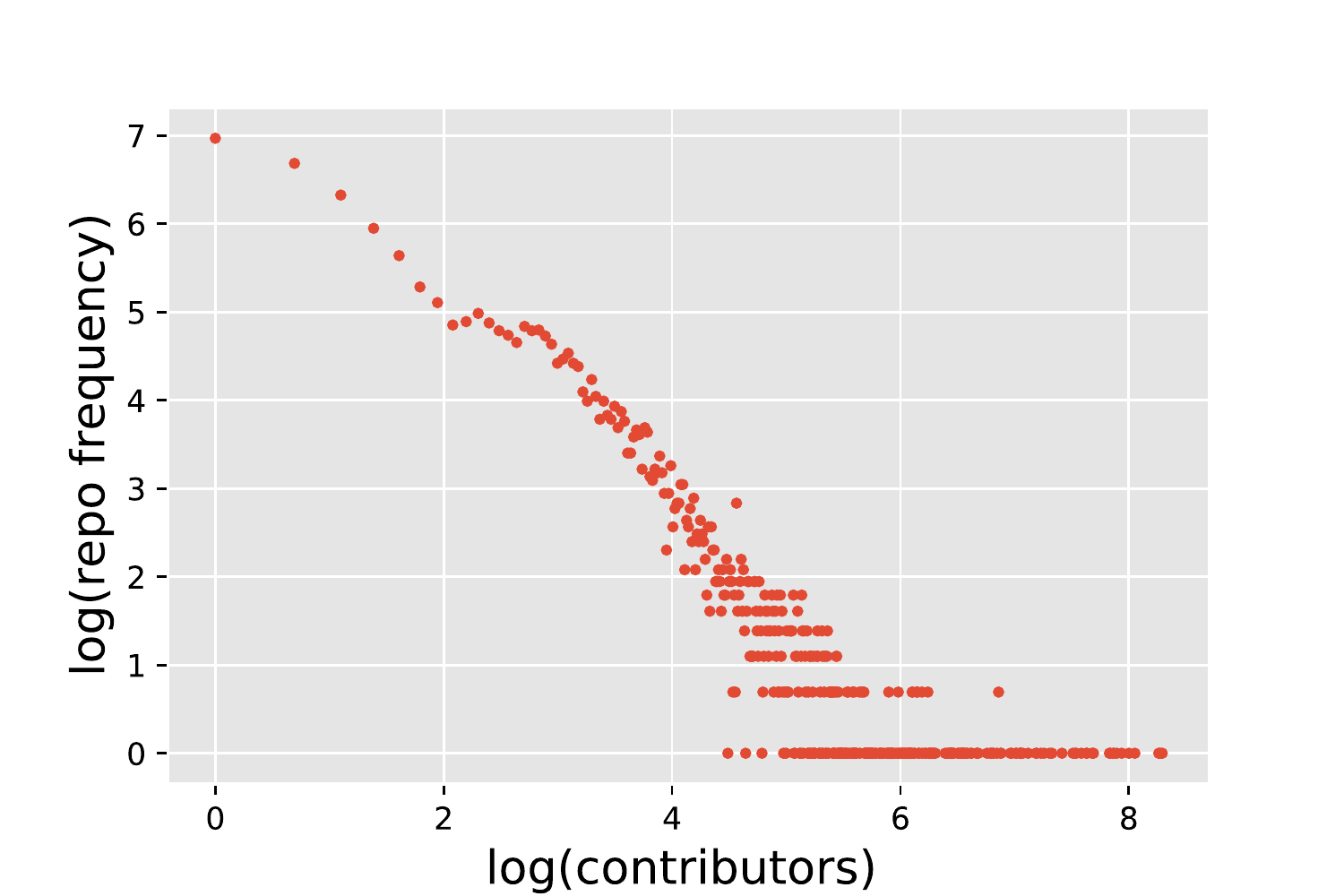}
        \label{power_law_repo_contributors}
    }
    \subfigure[Distribution of activities]{
        \includegraphics[width=0.24\textwidth]{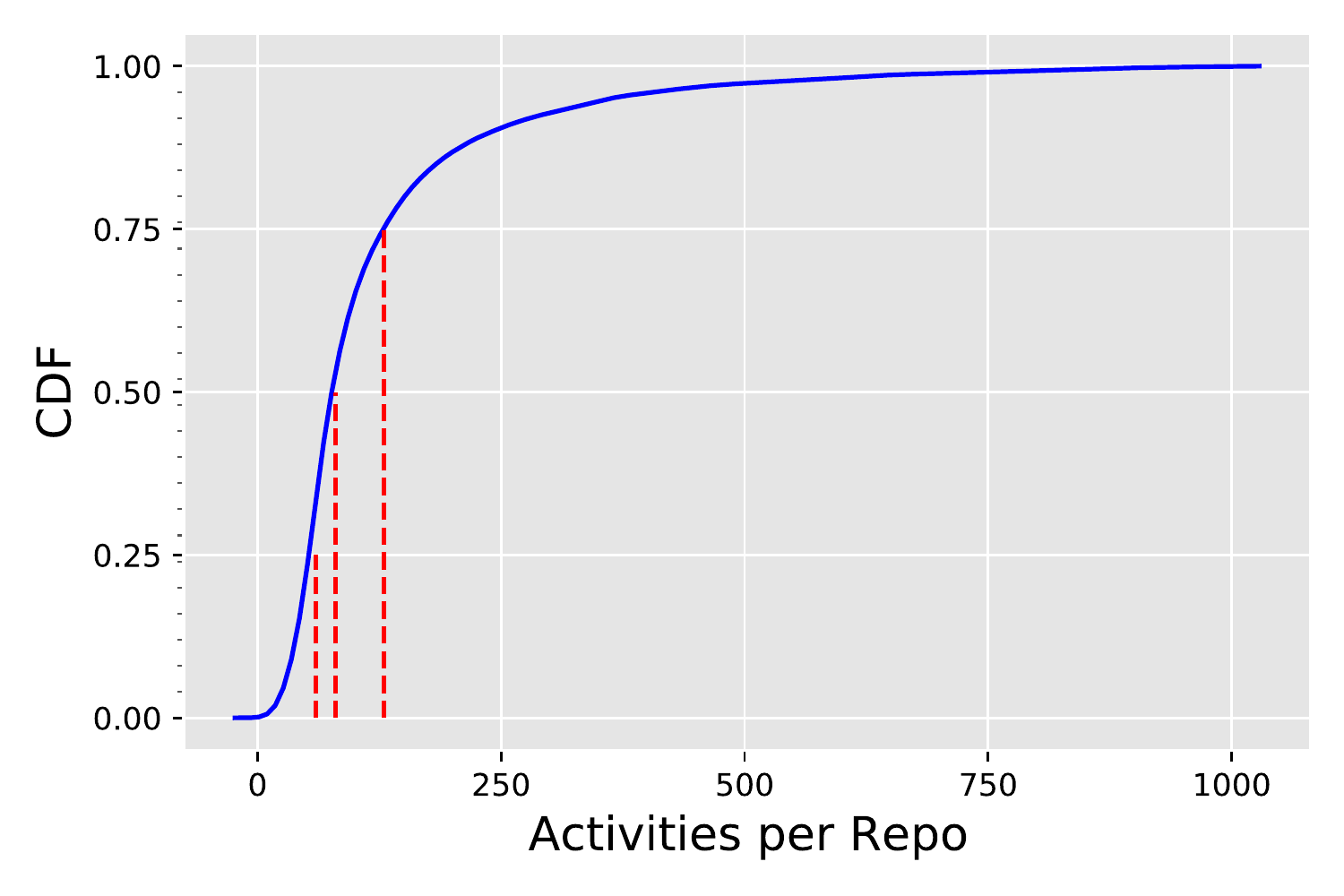}
        \label{repo_events_dist}
    }

    \setlength{\belowcaptionskip}{-11pt}
    \caption{Distribution of interactions on GitHub. (a) The Number of contributors per repository has a power-law-like distribution. (b) Distribution of activities over repositories has a long-tail; 75\% repos have less than 130 events.}
    \label{power_law_dist}
\end{figure}

\noindent\textbf{Power-law-like Distribution Implications.}  Similar to other social phenomena~\cite{easley2010networks}, we observe the interactions on GitHub follow a power-law-like distribution. Figure~\ref{power_law_repo_contributors} and Figure~\ref{repo_events_dist} show the distribution of the number of users and events per repository respectively. Following the power-law distribution, this means we observe only a few repos with many users and activities (a ``short head'') and many repos with only a few users and activities (a ``long tail''). Because of this, we propose a notion of clustering where repositories with similar sizes are clustered together for a fair evaluation. We separate our set of repositories into four groups using the amount of repository activity; we use these groups to evaluate the classifier performance. Figure~\ref{repo_events_dist} illustrates how we separate the repositories. Each vertical line corresponds to a separation point, which in turn corresponds to a quartile of repo-activity i.e., 25\%, 50\% and 75\% of repos have less than 60, 80 and 130 activities respectively. The artifact at the bottom-left of the plot~\ref{repo_events_dist} is explained by the fact that repos have a minimum number of events. Note that Watch events are excluded to focus only on development activities. We call each cluster as small, medium, medium-large and large reflecting the size of the repos in each cluster. Table \ref{repo_events_dist} reports the number of repos in each cluster separated by country. The majority class is ``China'' in the smallest cluster, and ``US'' in all others. This speaks to the relative size difference between the two repo groups. It is also noted that the number of repos associated with China in small cluster (409) is about twice as the number of repos in other Chinese clusters (medium: 229, medium-large: 229 and large: 258). We can relate this to the fact that many small repos originate from universities and such repos are more frequent in China (100/China vs 65/US university-based repos).

We first evaluate the effectiveness of our proposed repository representations using machine learning classifiers and a set of metrics including recall, precision, F1, and accuracy. We then set up an ablation study to understand the importance of the different representations through leave-one-out and decision stump approaches. Finally, we evaluate the impact of clustering on predicting country of a repository. 

\noindent\textbf{Performance Evaluation.} It is worth noting that we consider several types of classifiers as it is possible that some may perform better for this particular problem. They include Logistic Regression, Decision Tree, and Random Forest. These three represent linear, non-linear and ensemble methods respectively.We focus on the results of the logistic regression classifier which had the best performance for Recall, Precision, and F1. Detailed results, broken up by country and by quartile cluster are shown in Table~\ref{tb_LR}. Due to space limitations, we omit comparison with the other classifiers. 


The overall accuracy is 0.686, 0.747, 0.792 and 0.754 across clusters from small to large respectively. It indicates that repo representation has more power in distinguishing repos by country in large clusters than small ones. We speculate that this is because the social factors become more visible as developers interact more. 

 The F1 scores in US and China clusters are respectively (e.g. US / China): for small repos, 0.594/0.743, for medium repos, 0.791/0.681, for medium-large, 0.842/0.700 and for large, 0.793/0.700. The small cluster is the only one where China has a superior F1 score and it can be explained by the difference in number of repos in this cluster across US and China (342 vs 409).
 
In all cases, we are able to beat the majority class baseline wherein all testing samples receive the label as majority class, with the largest lift coming in the Large quartile. This provides an indication that more number of activities regarding a repository can lead to a better prediction. However, despite this insight, the quartile with the second-highest lift is the Small quartile. This suggests that the features we extract are also discerning when repos have less number of development activities. It is possible that different features are important for prediction in the Large quartile versus the Small. In the next section, we investigate the contribution of each feature through an ablation test.   

\begin{table}[tb!]
    \centering
    \caption{Repos by size separated by country.}
    \vspace{-2mm}
    \begin{tabular}{cccc}
        \toprule
         Size & US Repos & Chinese Repos & Total\\
         \midrule
         Small &342 & 409& 751 \\
         Medium & 364&229 & 593 \\
         Medium-Large &380 & 229& 609 \\
         Large &365 & 258& 623 \\
         \bottomrule
    \end{tabular}
    \label{tab:repocount}
\end{table}






\begin{table}[]
\scriptsize
\setlength\tabcolsep{4pt}
\caption{\label{tb_LR} Performance results of Logistic Regression classifier. A majority class baseline (``Majority'') is provided. }
\vspace{-3mm}
\begin{tabular}{lccccccccccc}
\toprule
\textbf{Cluster} & \multicolumn{3}{c}{\textbf{US-based}} & \multicolumn{3}{c}{\textbf{China-based}} & \multicolumn{5}{c}{\textbf{US and China} }\\
\cmidrule(lr){2-4}
\cmidrule(lr){5-7}
\cmidrule(lr){8-12}
& Prec & F1  & Prec & F1 & Prec & F1 & Acc & Majority\\
\toprule
\textbf{Small}     & 0.722 & 0.594 & 0.669 & 0.743  & 0.693 & 0.676  & 0.686 &  0.554 \\
\midrule
\textbf{Medium}    & 0.780 & 0.791 & 0.696 & 0.681  & 0.746 & 0.746  & 0.747 &  0.595 \\
\midrule
\textbf{Med.-lg.}  & 0.876 & 0.842 & 0.657 & 0.700  & 0.802 & 0.795  & 0.792 &  0.595 \\
\midrule
\textbf{Large}     & 0.739 & 0.793 & 0.779 & 0.700  & 0.757 & 0.750  & 0.754 &  0.551 \\
\bottomrule
\end{tabular}
\vspace{-1em}
\end{table}



\begin{table}[t]
\scriptsize
\setlength\tabcolsep{4pt}
\caption{\label{tb_ablation} Effectiveness of features (Accuracy). Results of an ablation test, where feature named in column is withheld from the model. A negative number means that the accuracy goes down without the feature i.e., the feature is effective.}
\vspace{-3mm}
\begin{tabular}{lcccccc}
\toprule
& Push & Fork & IssueCom & Issues & PullR & Create   \\
\toprule
\textbf{Small} & -0.005 & -0.005 &-0.005 & 0.000 & 0.000  &  0.000  \\
\midrule
\textbf{Medium} & 0.017  & 0.006 & 0.017 & 0.000 &  0.028 &   0.000 \\
\midrule
\textbf{Med-lrg} &-0.049 &-0.065 &-0.011 &-0.011 &-0.005 &  0.000   \\
\midrule
\textbf{Large} & 0.000  & 0.000 &  0.000 & 0.000  & 0.000   &  0.000 \\
\bottomrule
\toprule
& ReviewCom & Delete & Gollum & Member & Release & CommitCom  \\
\toprule
\textbf{Small} & 0.000 & 0.000 & 0.000 & 0.000 & 0.000 & 0.000    \\
\midrule
\textbf{Medium} &  0.028 & 0.028 & 0.028 & 0.000  & 0.000 &   0.000 \\
\midrule
\textbf{Med-lrg} & 0.000 & 0.000 &  0.000 &-0.005 & 0.000 &  0.000 \\
\midrule
\textbf{Large}&  0.000 &  0.000 &  0.000 &  0.000 &  0.000 &  0.000 \\
\bottomrule
\toprule
& Public & Jaccard & Leaders & ComLen & IAT  & Stars  \\
\toprule
\textbf{Small} & 0.000 &-0.005 & 0.022 & -0.053 & -0.005 & 0.017   \\
\midrule
\textbf{Medium} & 0.028 &  0.023 & 0.006 &-0.045 & 0.051 & 0.017 \\
\midrule
\textbf{Med-lrg} & 0.000 & -0.022 &-0.082 &-0.054 & -0.016 &  0.000 \\
\midrule
\textbf{Large} &  0.000 & 0.000 & -0.021 & -0.048 & 0.000 &  0.005 \\
\bottomrule
\toprule
& Forks & Open Issues & Topic$_{1}$ & Topic$_{2}$ & Seq-emb & All included \\
\toprule
\textbf{Small} & -0.009 & 0.014 & -0.005 &-0.005 &-0.009 & 0.000   \\
\midrule
\textbf{Medium} & 0.000  & 0.006  & 0.006 & 0.006 &  0.006 & 0.000 \\
\midrule
\textbf{Med-lrg} & -0.027 & -0.049 &-0.005 &-0.005 & -0.011 & 0.000  \\
\midrule
\textbf{Large} & 0.000 &  0.000 &  0.000 & 0.000 &-0.016 & 0.000 \\
\bottomrule
\end{tabular}
\end{table}
\begin{table}[t]
\scriptsize
\setlength\tabcolsep{4pt}
\caption{\label{tb_ablation_grp} Effectiveness of group of features. The accuracy of training the model \emph{only} on that group.}
\vspace{-3mm}
\begin{tabular}{lccccccc}
\toprule
 & Profile & Activity & Sequence & Profile   & Profile & Activity & All  \\
 &         &          &          & +Activity & +Sequence & +Sequence &   \\
\toprule
\textbf{Small} & 0.699 &  0.637&  0.593 &  0.677 &  0.681 & 0.637 &   0.686 \\
\midrule
\textbf{Medium} & 0.764 &  0.668 & 0.685 & 0.753 & 0.758 & 0.708 &   0.747 \\
\midrule
\textbf{Med-lrg} & 0.781 & {0.776} & 0.710 &  0.781 & 0.738 &  0.721 &  0.792 \\
\midrule
\textbf{Large} & 0.743 & 0.711 &  0.717 &  0.738 &  0.754 & 0.727 &  0.754 \\
\bottomrule
\end{tabular}
\end{table}

\noindent\textbf{Feature Importance (leave-one-out).}
We demonstrate significantly better than random performance on the task of differentiating by country. An important question is \emph{why} we're able to achieve this performance. We investigate this from the perspective of a feature ablation test, where we hold individual features from the model to study the effect this manipulation has on performance. The results are shown in Table~\ref{tb_ablation}. The values in this table are relative, computed as ``Accuracy without the feature'' -  ``Accuracy with all features'' (from Table~\ref{tb_LR}). Thus, a negative number represents an improvement in the presence of that feature and a decline in accuracy in the absence of that feature, in other words, the feature is effective. Altogether, Table~\ref{tb_ablation}   shows several interesting findings that we discuss below. 

\noindent\textit{Correlation with size of the repository.} The first observation we make about these results is that the effective features are vastly different based upon the size of the repo. For the medium-large and large repo set, we observe that the sequence embedding is one of the most effective features meaning 1.1 and 1.6 percent improvement in accuracy resp in the presence of these features. This indicates sequence embeddings contribute to the predictive power of the model while the classifier is better off without the sequence embedding features in medium cluster. This is an important indication that when number of development activities increases the sequence embedding can capture the style of the communication among developers more effectively.

\noindent\textit{Impact of number of leaders.} Another finding is the impact of number of leaders in medium-large and large repo set wherein we can observe an accuracy improvement of 8.2\% and 2.1\%, respectively, in the presence of this feature. This shows that the number of unique developers in larger repos  reflects the  structure of communication more effectively. 

\noindent\textit{Count-level features.} On the contrary, the repos belonging to the small cluster are differentiated the most by count-level features such as Push, Fork, and IssueCom. This is an important distinction, as it shows that when number of repo development activities are less, the model must rely on shallow count-based features, but as the amount of activities increases, the more sophisticated, complex features such as sequence embeddings become more pertinent and useful for differentiating between groups.

\noindent\textit{Neutral features.} From the $z$-test on percentage of activities, we found that distributions of features like ``ReleaseEvent'', ``GollumEvent'', ``CommitCommentEvent'', and ``PublicEvent'' are not different in repos associated with US and China significantly. This is mirrored in Table~\ref{tb_ablation} where withholding these features from the model does not impact the performance at all.

\noindent\textit{Length of Comments}. Another key observation is the impact of comment length in commit events across all clusters. The accuracy improves by 5.3\%, 4.5\%, 5.4\% and 4.8\% in small to large quartiles, respectively, in the presence of this feature. This echoes its power in distinguishing repo country.

Overall, in most cases there is not a noticeable drop in performance from omitting that particular feature. This indicates that, despite their disparate sources, there is a significant amount of redundancy in these features, and our finding regarding the amount of development activities is underscored here. When more development events are present, as in the large cluster, complex features like sequence embedding contribute more to predictive performance.


\noindent\textbf{Feature Importance (decision stump).} To delve deeper into the importance of features, we study the contribution of different \emph{groups} of features as profile, fingerprint activity and sequence embedding groups as discussed in Section 4. We do this by separating the features into groups and training the model with only that group. The results of this are shown in Table~\ref{tb_ablation_grp}.  The best overall model involves using the fingerprint activity features for differentiating country in the medium-large cluster (0.776) followed by large cluster (0.711). The similar pattern is noted in models built only on sequence embeddings for predicting the country in the medium-large (0.710) and large repos (0.717).

For repos with less activities--the small and medium--the profile features reign supreme. We observe that with these alone, we are able to achieve near-optimal performance on the Small clusters (small: 0.699 and medium: 0.764), with a decline in accuracy in the presence of the fingerprint activity and sequence embedding features. This also mirrors our findings from our leave-one-out study showing that when number of repo development activities are less, the model relies on shallow count-based features extracted from profile information more for differentiating between groups.  

Overall, grouping all repo representation factors outperforms exclusion of any one of them in medium-large and large groups. Profile-based features are more effective in small and medium clusters and their combination with activity and sequence based features actually hurts the performance of profile features alone.


\noindent\textbf{Impact of Clustering.} Most similar approaches in the past have focused primarily on cases where data like the activity data is plentiful. In this paper, we do not limit our analysis to the repositories with more than certain number of activities, but expand our analysis across repos with various sizes by breaking out them into different clusters. To evaluate the impact of clustering, we also measure the performance when the model is developed over a single group of repos collectively.  The \textit{single group performance}/ \textit{average of performance across four clusters} is precision: 0.7148/ 0.7495, recall: 0.7141/ 0.7442, f1: 0.7144/ 0.7417 and accuracy: 0.7141/ 0.7447 showing more than three percent improvement when repo representations are evaluated within different cluster groups. An outcome of this work is the impact of clustering on predicting repo country.

\section{6. Country versus Corporation Associations} \label{Corporation}

Finally, we use a case study to explore how development activities on GitHub differ between different organizational groups by looking at repositories associated with different companies versus different countries.

Throughout the course of this paper, we have been interested in identifying differences in software development based on a repository's country as country is a form of organization separate from company that may still contribute to differences in development. Individuals have associations with multiple organizations, country being one of the larger and more nebulous. We have demonstrated that there are verifiable differences between repositories associated with different countries. We are curious how a similar approach may perform when looking at more traditional organizational structures like corporations. In this section, we investigate the extent to which company may influence the development of the repository. 
We do this through the lens of a case study, identifying prominent corporations and their major repositories on GitHub.  We collect 223,444 events for 27  repositories across six different companies. An overview of the repositories can be seen in Table~\ref{tab_cs_repos}.

\begin{figure}[tb!]
    \centering
    \hspace*{0.5cm} 
    \includegraphics[width=0.9\linewidth]{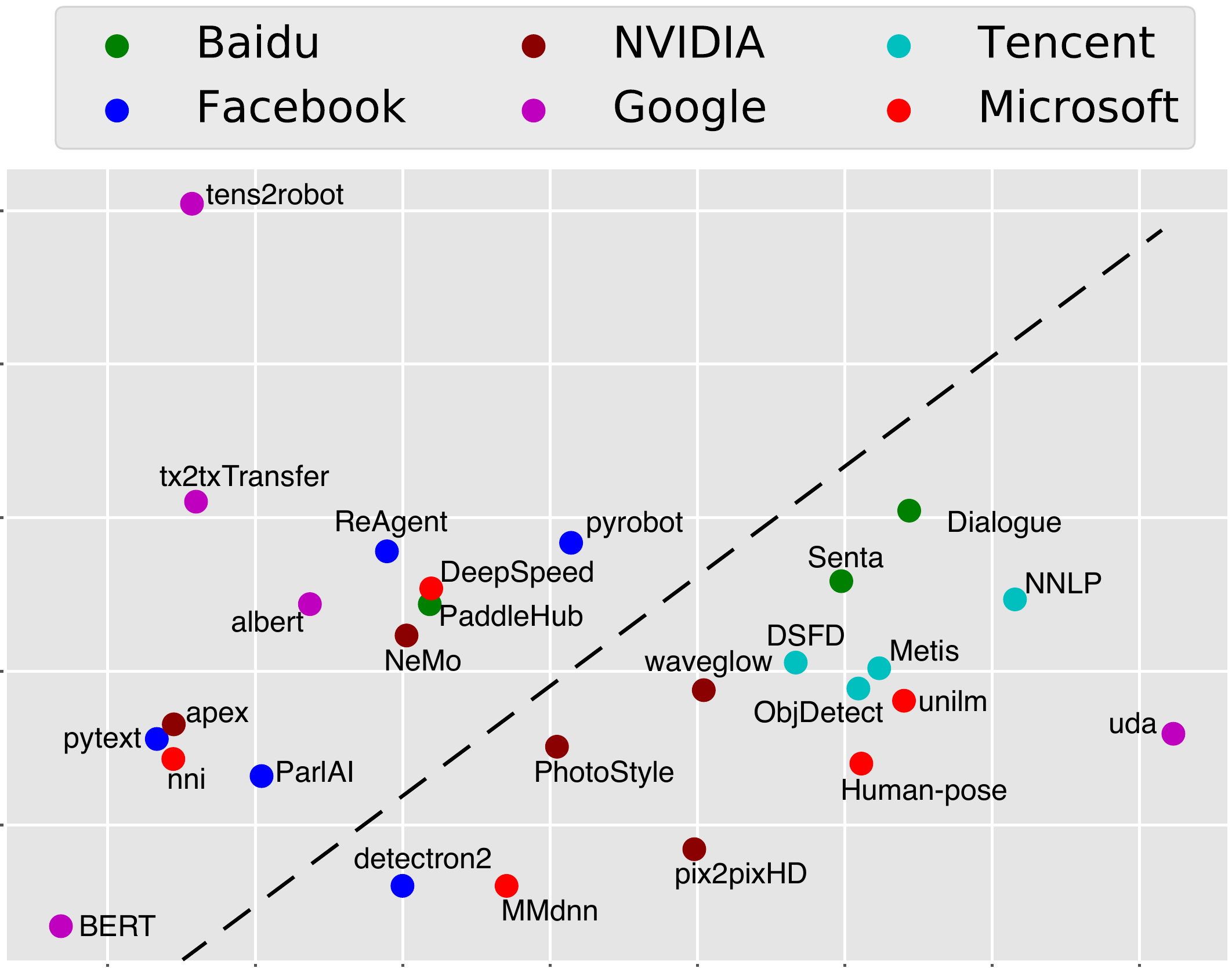}
    \caption{PCA 2D projection of repositories across different companies. Colors denote the company and the repository names are shown next to each data point. The dotted line indicates country association where repos below and above the line are almost entirely from China and US resp.}
    \label{fig_case_study_pca}
    \vspace{-2em}
\end{figure}

Table~\ref{tab_cs_repos} shows how the repositories are labeled by the k-nearest neighbor approach; each repository receives the label that is carried by majority of its neighbors based on Euclidean distance metrics. We see that NVIDIA, Tencent, and Baidu are often confused for each other by the model. Structurally, this is confirmed by PCA (Figure~\ref{fig_case_study_pca}) in the positioning of the repos. However, we must note that NVIDIA has connections to China, with a large portion of its operation based there. The algorithm is potentially picking up the regional culture when making its classification. Cultural effects can be seen again in the unilm repository. While this particular Microsoft repository was mapped to Tencent, it is important to note that the group that developed unilm was based out of the Beijing office of Microsoft Research. We verified this by looking at the authors' location from their LinkedIn and Microsoft pages. Similarly, human-pose-estimation.pytorch, which was labeled as NVIDIA, was developed by Microsoft Research Asia~\cite{xiao2018simple}. In these cases, we see the effect of country-cultural effects overshadows that of company.

\begin{table}[t]
\caption{Corporate repositories used in our case study and their predicted labels using KNN algorithm.}
    \centering
    \begin{tabular}{lcc}
    \toprule
    Organization & Repository name & Label by KNN \\
    \midrule
Facebook &  ParlAI    & Facebook\\
Facebook  & ReAgent  & Facebook \\
Facebook  & pytext & Facebook \\
Facebook  & pyrobot & Baidu\\
Facebook  & detectron2 & NVIDIA\\
\midrule

NVIDIA           &  waveglow & NVIDIA\\
NVIDIA           &  NeMo & NVIDIA\\
NVIDIA          &  apex & NVIDIA\\
NVIDIA          &  FastPhotoStyle & Tencent\\
NVIDIA          &  pix2pixHD & Microsoft\\
\midrule

Tencent           &  ObjectDetection- & Tencent\\
 & OneStageDet & \\
Tencent            &  Metis & NVIDIA\\
Tencent           &  FaceDetection-DSFD & NVIDIA\\
Tencent            &  NeuralNLP- & NVIDIA\\
 & NeuralClassifier & \\
\midrule

Baidu           &  Dialogue & Baidu\\
Baidu           & Senta & Tencent\\
Baidu           &  PaddleHub & NVIDIA\\
\midrule

Google           &  tensor2robot &  Google \\
Google           & Albert & Facebook\\
Google           &  text-to-text- &  Facebook\\
                         & transfer-transformer & \\
Google           &  BERT & NVIDIA\\
Google           &  uda & Tencent\\
\midrule

Microsoft          &  DeepSpeed & Facebook\\
Microsoft          & MMdnn & Facebook\\
Microsoft          &  nni & Facebook\\
Microsoft          &  human-pose- & NVIDIA\\
                   & estimation.pytorch & \\
Microsoft          &  unilm & Tencent\\

    \bottomrule
    \end{tabular}
    \label{tab_cs_repos}
\end{table}

We do also see instances where company overrides country. DeepSpeed, developed by Microsoft, is misclassified as a Facebook repository. An important point is that the DeepSpeed library is based upon PyTorch, a popular library from Facebook. DeepSpeed's similarity to Facebook repositories may be explained by Conway's law, with the individuals who developed these similar libraries potentially using similar communication structures. Figure~\ref{fig_case_study_pca} shows that Facebook  repositories fall in the neighborhood of DeepSpeed.  

While we have found both cases of country and company-level artifacts in the data, Figure~\ref{fig_case_study_pca} shows a larger pattern in the differences between countries. The dotted line on the graphic was drawn by us and indicates country association. The repositories below this line are almost entirely from China and those above almost entirely from the US. This suggests stronger separation in feature space with respect to the attributes associated with location. In other words, country drives the majority of the difference.


\section {7. Conclusion and Discussion}
In this work we study the extent to which software development differs depending on the country developers associate with. We collect a novel dataset containing repositories from different countries, collecting the code, its changes, and its event history. We extensively study these repos through the lens of several sequence, activity, and profile measures. We find that there are statistically significant differences in the activity patterns between repos associated with China and the USA. This echoes a cultural twist on Conway's law, that not only the company but also country can influence the design of code. Using off-the-shelf classification techniques built with the measures we designed, we find that we are able to classify the repositories with up to 79.2\% accuracy on certain sets of features. Lastly, we conduct a case study where we investigate the influence of country-level and company-level differences. We find that both may contribute to software development but that country appears to more frequently explain differences in software development features between repositories. 

A unique outcome of this work is the impact of clustering on predicting the country of a repository. Most similar approaches in the past have focused on cases where data like our activity sequences is plentiful. While we removed repositories which had little to no activity, we retained repositories where activity was present but the activity sequences were less prolific in comparison to some of the larger repos. By treating these repos separately, we were able to show the value of different feature representations for predicting country depending on the size of the repository. Another advantage of this work is that the proposed repo representation methodology relies on both hand-crafted and end-to-end features. In ongoing efforts, we aim to expand our sequence embedding models by incorporating contextual information such as timestamp and developers IDs to capture the underlying communication structure of the code development activities more effectively.   

There are several limitations to this work. \textit{First,} GitHub is a constrained resource and our review is automatically biased by the factors that determine whether a developer will utilize GitHub. \textit{Second,} the labels we utilize are based purely on what users claim in their GitHub profiles and we have focused on just two countries for this initial exploration. Future work should focus on additional methods for identifying country and incorporate additional countries. \textit{Third,} this work focuses only on development activities while the source code itself also can convey the way contributors develop. Analyzing source code through the lens of Conway's law is an interesting direction for future. 

Again, dataset and codes will be released on a public site. Due to space limitations, we add the Reproducibility section where we discuss the hyper-parameters later.


\bibliography{refs}
\bibliographystyle{aaai}

\end{document}